\documentclass[11pt]{article}
\usepackage[a4paper,bindingoffset=0.2in,%
            left=1.2in,right=1.2in,top=1.5in,bottom=1.5in,%
            footskip=.25in]{geometry}

\usepackage[T1]{fontenc}
\usepackage[utf8]{inputenc}
\usepackage{authblk}
\usepackage[numbers]{natbib}

\usepackage{amsthm,amsmath}
\usepackage{amsfonts}

\theoremstyle{definition}

\usepackage{footnote}
\usepackage{verbdef}
\usepackage{booktabs, makecell, tabularx}
\usepackage{float}

\makeatletter
\newtheorem*{rep@theorem}{\rep@title}
\newcommand{\newreptheorem}[2]{%
\newenvironment{rep#1}[1]{%
 \def\rep@title{#2 \ref{##1}}%
 \begin{rep@theorem}}%
 {\end{rep@theorem}}}
\makeatother

\makeatletter
\newcommand\footnoteref[1]{\protected@xdef\@thefnmark{\ref{#1}}\@footnotemark}
\makeatother

\newreptheorem{theorem}{Theorem}
\newreptheorem{lemma}{Lemma}
\usepackage{graphicx}
\graphicspath{{images/}{../images/}}
\usepackage{booktabs}

\usepackage{bbm}
\usepackage{amsmath}
\usepackage{xcolor}

\usepackage{algorithm}
\usepackage{algorithmic}

\usepackage{newfloat}
\usepackage{listings}
\floatstyle{ruled}
\newfloat{listing}{tb}{lst}{}
\floatname{listing}{Listing}

	\newcommand{\blind}{0}
	
	\usepackage{amsmath}
	\usepackage{graphicx}
	\usepackage{enumerate}
	\usepackage{url} % not crucial - just used below for the URL
\begin{document}
		
		\def\spacingset#1{\renewcommand{\baselinestretch}%
			{#1}\small\normalsize} \spacingset{1}

		\if0\blind
		{
			
			\title{Node Attribute Prediction on Multilayer Networks with \\Weighted and Directed Edges}	
			
	\author[1]{Yiguang Zhang}
	\author[2]{Kristen Altenburger}
	\author[2]{Poppy Zhang}
	\author[2]{Tsutomu Okano}
	\author[2]{Shawndra Hill}
	\affil[1]{Columbia University}
	\affil[2]{Meta Central Applied Science}

%			\author{Yiguang Zhang$^a$, Jessy Xinyi Han$^b$, Ilica Mahajan$^a$, \\
%			Priyanjana Bengani$^a$, and Augustin Chaintreau$^a$ \\
%			$^a$ Columbia University \\
%             $^b$ Massachusetts Institute of Technology }
			\date{}
			\maketitle
		} \fi
		
		\if1\blind
		{

            \title{\bf \emph{IISE Transactions} \LaTeX \ Template}
			\author{Author information is purposely removed for double-blind review}
			
\bigskip
			\bigskip
			\bigskip
			\begin{center}
				{\LARGE\bf \emph{IISE Transactions} \LaTeX \ Template}
			\end{center}
			\medskip
		} \fi
		\bigskip

\begin{abstract}

% Edits from Nico:
% \begin{itemize}
%     \item for protein network - highlight results more in beginning about single-layer vs. multi-layer
%         \item TODO redo wilcoxon tests
%     \item TODO update notebooks that read in summary
%     \item TODO fix github repo
%   \item in github make a note that we don't include + constants
% \end{itemize}

With the rapid development of digital platforms, users can now interact in endless ways from writing business reviews and comments to sharing information with their friends and followers. As a result, organizations have numerous digital social networks available for graph learning problems with little guidance on how to select the right graph or how to combine multiple edge types. In this paper, we first describe the types of user-to-user networks available across the Facebook (FB) and Instagram (IG) platforms. We observe minimal edge overlap between these networks, indicating users are exhibiting different behaviors and interaction patterns between platforms. We then compare predictive performance metrics across various node attribute prediction tasks for an ads click prediction task on Facebook and for a publicly available dataset from the Open Graph Benchmark. We adapt an existing node attribute prediction method for binary prediction, LINK-Naive Bayes, to account for both edge direction and weights on single-layer networks. We observe meaningful predictive performance gains when incorporating edge direction and weight. We then introduce an approach called MultiLayerLINK-NaiveBayes that can combine multiple network layers during training and observe superior performance over the single-layer results. Ultimately, whether edge direction, edge weights, and multi-layers are practically useful will depend on the particular setting. Our approach enables practitioners to quickly combine multiple layers and additional edge information such as direction or weight. 
\end{abstract}

\section{Introduction}
Inferring user attributes has a rich history in the graph mining literature including predicting demographic attributes~\cite{dong2017user,brea2014harnessing}, marketing behavior~\cite{hill2006network,goel2014predicting,zubcsek2017predicting}, integrity outcomes~\cite{chakrabarti2014joint,fakhraei2015collective,wang2017gang} and more. The typical within-network setting includes known ground-truth labels for a subset of users and a single network on how all users are connected. The goal is then to infer the missing labels or attributes for the remaining nodes~\cite{neville2000iterative,welling2016semi,perozzi2014deepwalk,sen2008collective,altenburger2021node}. The related across-layer task trains on a set of labeled nodes and edge type and predicts for the same node set for a different edge type. Examples include making attribute predictions over time~\cite{henderson2011s}. A recent review paper has also unified the various types of graph representation learning methods~\cite{chami2020machine}. We focus on the within-network task in this paper. % but will consider how to combine multiple layers during training.

Feature representations for node attribute prediction problems can generally be classified by whether the features are identity-independent (dependent) and label-independent (dependent)~\cite{altenburger2021node}. Identity-dependence captures whether a feature depends on the identity labels of the node $i$=1,2,...,$n$. Label-dependence~\cite{gallagher2008leveraging} captures whether the feature depends on the attribute label. Node embedding feature representations are a popular class of methods for representing users via a low-dimensional feature. Approaches like node2vec~\cite{grover2016node2vec}, DeepWalk~\cite{perozzi2014deepwalk}, and LINE~\cite{tang2015line} are all identity-dependent and label-independent. LINK~\cite{zheleva2009join} is another promising identity-dependent and label-independent feature representation where the feature is the row in the adjacency matrix.  

While many within-network tasks train on a single network or single edge relationship~\cite{henderson2012rolx,henderson2011s,altenburger2021node,lim2021new}, real-world node prediction settings have numerous user-to-user networks to consider for a given prediction task. It is not clear which is the optimal network to select for a given problem. For example, Facebook (FB) users can be connected by a friendship edge or interaction edges including comments, reactions, and more. These same users can also follow one another on Instagram (IG) and interact. On other social network platforms like Twitter, users can be represented by follower relationships or interaction edges~\cite{huberman2008social}. These types of networks are considered multi-layer networks~\cite{kivela2014multilayer} where the two aspects are the platform (FB/IG) and interaction or friendship/follower edge type. 

With the rise in multi-relational edge types in networks, recent methods propose new ways to incorporate multi-layer edges. Embedding approaches can additionally incorporate multi-relational edges into the embedding framework~\cite{feng2019marine,bordes2013translating,lerer2019pytorch,zitnik2017predicting}. Graph Neural Network (GNN) approaches %https://arxiv.org/pdf/2109.10119.pdf
include approaches like heterogeneous graph transformers~\cite{hu2020heterogeneous} which can incorporate different node and edge types. Meanwhile, other promising feature representations like LINK~\cite{zheleva2009join}, which create a large sparse feature vector from the row of an adjacency matrix, do not account for different edge types. 

In this paper, we extend the LINK approach~\cite{zheleva2009join} to weighted and directed edges on single-layer networks and then extend the method to multi-relational edges. This paper follows other recent approaches for leveraging multilayers such as modeling multilayer interdependence~\cite{aguiar2022factor} or extracting $k$-cores from multilayers~\cite{hashemi2022firmcore}. We refer the interested reader to reviews~\cite{boccaletti2014structure,kivela2014multilayer} on multilayer networks. Our work is also related to the ``graph sanitation'' task~\cite{xu2022graph}, which learns the optimal graph representation from a single graph. In our setting, our goal is to learn the relative value of different types of edges based on weight, direction, or multi-layers.

We first introduce and measure the similarity across the various user-to-user networks for the Facebook and Instagram platforms in the ``Analysis of Multilayer Networks'' section. We find these different user networks have minimal edge overlap, indicating users are interacting with different sets of users across platforms. In the ``Node Attribute Prediction with Single-layer Networks'' section, we apply LINK-Naive Bayes to a business prediction task and to a protein-protein network that's publicly available. We introduce a new concept of ``edge binning'' that enables us to distinguish between different types of edges. We observe %heterogeneity in the predictive performance across the undirected/unweighted friendship network and the directed/weighted interaction network representations. 
improved predictive performance when incorporating edge direction and/or edge weight for both tasks. Interestingly for the protein-protein network, we observe an average ROC-AUC performance of 0.51 on the undirected/unweighted network but observe improved performance ranging from 0.57-0.70 for the different versions incorporating edge weight information for this network.
In the ``Node Attribute Prediction with Multi-layer Networks'' section, we introduce a method called MultiLayerLINK-NaiveBayes which combines multiple layers for network prediction and observe additional predictive performance gains. We conclude in the ``Discussion and Future Work'' section with a discussion of our findings and on future research directions in the field of node attribute prediction.

\textit{Broader Perspectives and Ethical Considerations: }This research was reviewed and approved by an internal review board-like organization in industry. We limit all analysis to users who have explicitly linked their Facebook and Instagram accounts. All data was de-identified and analyzed in aggregate. Node attribute prediction challenges in particular and machine learning on graphs in general have become commonplace across organizations with rich social network data. Multilayer networks provide a natural language for analyzing different user-level networks and for framing various prediction tasks. New methods to improve how to optimally combine or select networks for a prediction task have tremendous potential for helping organizations improve the user experience. {\color{black}Beyond what is presented here, we believe there is room for improved methods that evaluate the representativeness of node labels for prediction tasks and adequately accounts for any label bias.}  

\section{Analysis of Multilayer Networks}
\label{sec:analysis}

A multilayer network~\cite{kivela2014multilayer,boccaletti2014structure,de2013mathematical} consists of a node set $V$ of $N$ nodes that can belong to different layers $L$ such as a Facebook friendship edge or a comment edge between users. A multiplex network is a type of multilayer network that is limited to the same set of nodes across different network layers. While the approaches presented in this paper are applicable to the general multilayer set-up, we illustrate the methodology on multiplex networks.

While prior work analyzed the social structure on the FB friendship graph~\cite{ugander2011anatomy}, no work to our knowledge has done an empirical comparison of different network representations of users across different interactions. We can view these different representations as a multilayer network~\cite{kivela2014multilayer} and compare users (nodes) between layers. For each platform (FB and IG), we pull data for 2-weeks in June 2022 for active U.S users over 18-years old with at least 20 Facebook friends, and construct user-to-user interaction networks (FB/IG), friend and follow networks (FB/IG). Interaction networks are directed and weighted and can include numerous interaction types including comments, reshares, likes, comment likes, and mentions. We direct the interaction from the source user creating the interaction to the target user receiving the interaction. We limit all analysis to users who have explicitly linked their FB and IG accounts. All data was de-identified and analyzed in aggregate. %Researchers did not view any individual-level data. %We limit all analysis to users who ``hard-link'' their account or give permission to specifically link different accounts.

On IG, the out-degree distribution across all interactions has more weight on small degrees, while in-degree distribution is more heavy-tailed. That is, there is a larger proportion of users receiving many interactions than the proportion of users sending many interactions. We also observe that different interactions have different patterns with respect to who people interact with: direct reshares and mentions happen mainly between users who follow each other (``reciprocal follows''). %who know each other such as friends and family. 
Comment, overall likes, and likes on comments also happens more among reciprocal follows. %non-friends and family relationships. 
On FB, in terms of degree distributions, unlike IG, for different interactions, the out-degrees do not always have more weight on small degrees except for comment and reshares. Users primarily interact with those with reciprocal follows. %friends and family.
When comparing networks across IG and FB, the edge overlap is small indicating that a user's connections on IG are different than those on FB. We find small correlations between structural properties like degree. In total, these empirical observations indicate that different user network layers may provide different values for node prediction tasks. 

\section{Single-layer Networks}
\label{sec:prediction}
While existing node embedding approaches can incorporate multi-relational information~\cite{feng2019marine,bordes2013translating,lerer2019pytorch}, other existing feature representations do not. LINK-logistic regression~\cite{zheleva2009join} is an interpretable node prediction approach that represents a node by the row of the graph adjacency matrix. This large, sparse feature vector can then be utilized by a regularized logistic regression model for binary node prediction tasks. This method can also perform optimally with or without homophily~\cite{zheleva2009join,altenburger2018monophily}.  For this paper, given the flexibility of LINK to work in a variety of settings and given its prior high predictive performance, we adapt the LINK-Naive Bayes approach~\cite{altenburger2018monophily} to account for both edge direction and weights. We note that LINKX is a similar approach~\cite{lim2021new} that embeds the node features and adjacency matrix separately via an MLP. The focus of this work is on LINK-Naive Bayes where we have only the adjacency matrix as an input signal. While LINK refers to using the rows of the adjacency matrix as features, the scope of this paper is on node attribute prediction and not link prediction.

\begin{figure}%[h]
    \centering
    \includegraphics[width=0.7\textwidth]{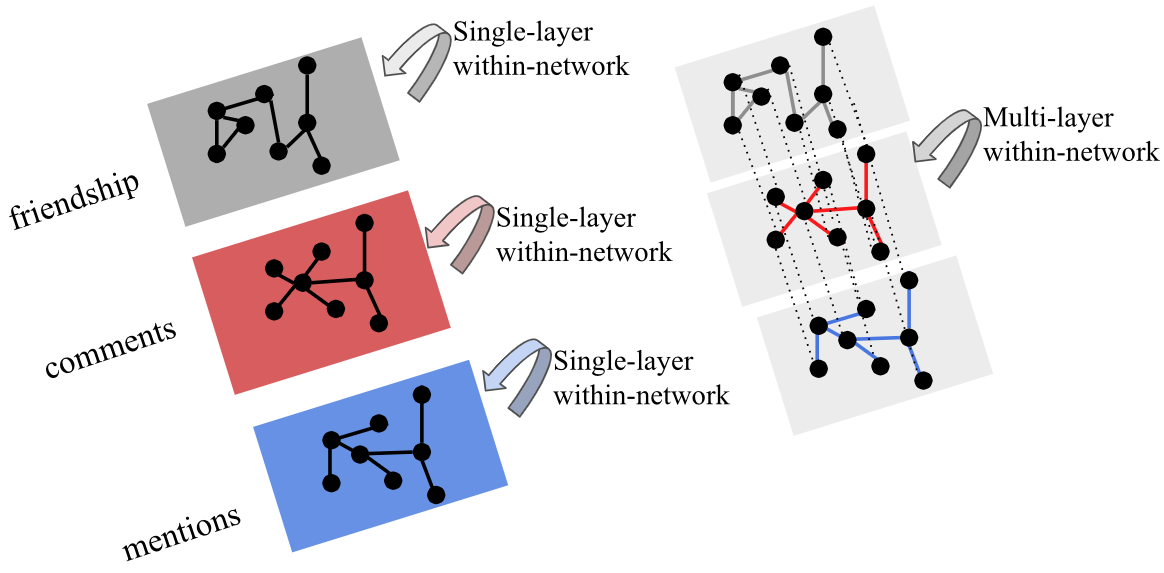}
    \caption{
    (left) We illustrate a single-layer network prediction problem across different edge types. Each layer represents a particular relation or interaction type and, we compare predictive performance of each layer. (right) We illustrate how multi-layer prediction is done with respect to different edge types. }
    \label{fig:single_layer_viz}
    \vspace{-0.6cm}
\end{figure}

\subsection{Undirected and Unweighted Edges}
We first review the derivation of LINK-Naive Bayes for a single layer in an undirected, unweighted network $G = (V,E)$, where each $v\in V$ represents a user, and $(u,v) \in E$ if and only if $u$ and $v$ interact with each other. We show in Figure~\ref{fig:single_layer_viz} (left) a schematic of the within-layer task. Each rectangle represents a network for a particular edge type or layer. The prediction task, as will be explained, will be done within each network such as only within the friendship network. 

Suppose we have a binary label prediction task, where each node has a label being either $+1$ or $-1$. For a target node $u$, the task is to predict its label using known labels from neighbors of $u$ in the network layer. Particularly, for each node $i \in G$, we define $x_i:=\mathbbm{1}_{\{(i,u) \in E(G)\}}$ as the indicator of $i$ being a neighbor of $u$, and a random variable $X \in \{0,1\}^N$, where $N$ is the total number of nodes in the network. We  review the Naive Bayes classification rule via the likelihood ratio and by making a conditional independence assumption:

\begin{align*}
LR(x) &= \frac{P(y_{train}=+1)\cdot P(X | y_{train}=+1)}{P(y_{train}=-1)\cdot P(X | y_{train}=-1)}\nonumber \\
&= \frac{P(y_{train}=+1)\cdot \prod_{i=1}^{N}P(X_i = x_i | y_{train}=+1)}{P(y_{train}=-1)\cdot \prod_{i=1}^{N}P(X_i = x_i | y_{train}=-1)}\nonumber \\
&= \frac{P(y_{train}=+1)}{P(y_{train}=-1)} \cdot \prod_{i:x_i = 0} \frac{P(X_i = x_i | y_{train}=+1)}{P(X_i = x_i | y_{train}=-1)}\nonumber \cdot \prod_{i:x_i = 1} \frac{P(X_i = x_i | y_{train}=+1)}{P(X_i = x_i | y_{train}=-1)}
\end{align*}

\noindent Note that by setting $n_{+1}$ ($n_{-1}$) as the number of $+1$-labeled ($-1$-labeled) nodes in the training sample and $d_{i,+1}$ ($d_{i,-1}$) as node $i$'s degree with $+1$ ($-1$) nodes in the training sample, we have the following empirical estimations:

\begin{align*}
\hat{P}(y_{train}=+1) = \frac{n_{+1}}{n_{+1} + n_{-1}},\\
\hat{P}(y_{train}=-1) = \frac{n_{-1}}{n_{+1} + n_{-1}}; \\
\hat{P}(X_i = 1 | y_{train}=+1) = \frac{d_{i,+1} + 1}{n_{+1} + 2}, \\
\hat{P}(X_i = 1 | y_{train}=-1) = \frac{d_{i,-1} + 1}{n_{-1} + 2}; \\
\hat{P}(X_i = 0 | y_{train}=+1) = \frac{n_{+1} - d_{i,+1} + 1}{n_{+1}+2},\\
\hat{P}(X_i = 0 | y_{train}=-1) = \frac{n_{-1} -d_{i,-1} + 1}{n_{-1}+2}.
\end{align*}

\noindent Assuming that for sparse networks that $n_{+1}, n_{-1} >> d_{i,+1}, d_{i,-1}$, the final derivation for the log of the likelihood ratio can be written as follows~\cite{altenburger2018monophily}:

\begin{eqnarray}
\label{log-LR-prev}
log(LR(x)) \approx \underbrace{\sum_{i=1}^{N} x_i \cdot \underbrace{log\left[ \frac{d_{i,+1}+1}{d_{i,-1}+1} \cdot \frac{n_{-1}+2}{n_{+1}+2} \right]}_\text{one-hop aggregation}}_\text{two-hop aggregation}.
\end{eqnarray}

\subsection{Directed and Weighted Edges}
As we observed when comparing user-to-user networks in the ``Analysis of Multilayer Networks'' section, edge direction and weights can contain different information across different interaction networks. To better use network signals from directed and weighted networks, we propose two extensions of LINK-Naive Bayes to incorporate edge direction and weight.

\subsubsection{Create Edge Bins: }
\begin{figure}[h]
    \centering
    \includegraphics[width=0.4\textwidth]{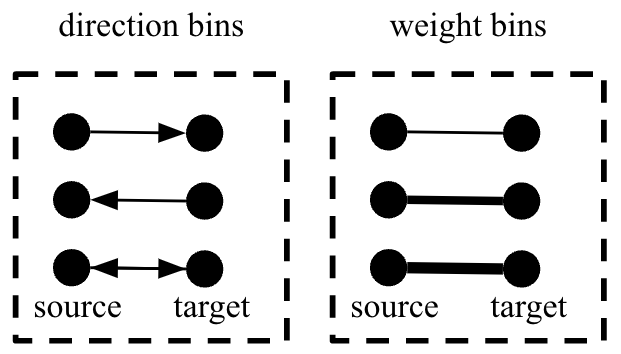}
    \caption{
    We illustrate binning edges. (left) We show 4 categories for binning edges based on whether a source node connects with a target node, a target node connects with a source node, both source and target connect, or neither. (right) We show 4 categories for binning edges based on the strength of the connection between the source and target node. }
    \label{fig:edge_binning}
\end{figure}

One central idea for our approach is to create ``bins'' for edges. We will describe the different ways to create edge bins and will reference this set-up throughout the rest of the paper. An edge bin is a way to place edges into mutually exclusive categories based on edge weight and/or edge direction. As shown in Figure~\ref{fig:edge_binning}, we illustrate edge bins based on direction (left) and edge bins based on weight (right). This means that instead of treating all edges equivalently as in the current LINK-Naive Bayes approach, edge bins will be a way for us to distinguish between different types of edges. We can also combine different edge bins to create finer bins such as combining direction $\times$ weight bins. 

\subsubsection{(Version I) Learning Weights for Edge Bins:}
We change the previously introduced observed features $x_i \in \{0,1\}$ and the random variable $X \in \{0,1\}^N$ to reflect edge weights and/or directions. Specifically, we first divide the edges into $|W|$ bins by pre-selected metrics. For example, as introduced in the last section, edges with different weights are divided into bins by their values, and directed edges are divided into bins by whether the edge is an outgoing one from the target node, an incoming one, or both. We denote the set of bins as $\{Bin_w, w \in W\}$, where $W$ is the set of bins. We denote $\mathbbm{1}_{x_i = w}$ if the edge between the target node $j$ and source node $i$ belongs to $Bin_w$. 

An easy extension to test is optimizing the weights on different edge bins for two-hop aggregations when aggregating scores among a source node's connections (right-hand side of Equation \ref{log-LR-prev}). Specifically, let $\lambda_w$ be the weight of $Bin_w$ and denote $\vec{\lambda}$ the vector of $\lambda_w$'s: $\vec{\lambda} = (\lambda_w)_{w = 1, \ldots, |W|}$. Denoting $\alpha$ as the intercept term in Logistic regression, we can write the classifier's score function $f(.)$ as the following: 

\begin{align}
\label{log-LR-weighted-v1}
&f(\alpha, \vec{\lambda}, x) = \alpha + \sum_{w = 1}^{|W|} \lambda_w \underbrace{\sum_{i=1}^N \mathbbm{1}_{x_i = w} \cdot \underbrace{log\left[ \frac{d_{i,+1}+1}{d_{i,-1}+1} \cdot \frac{n_{-1}+2}{n_{+1}+2} \right]}_\text{one-hop aggregation}}_\text{two-hop aggregation by edge bin $w$}
\end{align}
We can then estimate $\alpha$ and 
$\vec{\lambda}$ with logistic regression utilizing a validation split, which is minimizing the log loss for: 

% to minimize the Log Loss: 
\begin{align*}
\label{NB_loss}
    l := \sum_{(X, y) \in {\text{validation data}}} -\left(\frac{y+1}{2}\right) \log\left(\frac{e^{f(\alpha, \vec{\lambda}, X)}}{1+e^{f(\alpha, \vec{\lambda}, X)}}\right) 
         - \left(\frac{y-1}{2}\right) \log\left(\frac{1}{1+e^{f(\alpha, \vec{\lambda}, X)}}\right).
\end{align*}

\noindent Note that we can regard Equation \ref{log-LR-weighted-v1} as a more general form of Equation \ref{log-LR-prev} as it recovers Equation \ref{log-LR-prev} in the case that, there are only two bins with $Bin_1 = \{x_i : x_i = 1\}, Bin_2 = \{x_i : x_i = 0\}$, and  $\lambda_1 = 1, \lambda_2 = 0, \alpha = C_1$. The terms $d_{i,+1}$ ($d_{i,-1})$ still refer to the undirected/unweighted degree counts a source node has with other nodes labeled $+1$ ($-1$).

\subsubsection{(Version II/II$^{\ast}$) Classifier within each Edge Bin: }
An alternative approach is to apply a Naive Bayes classifier within each edge bin. Here, we further denote $d_{i,+1,w}$ ($d_{i,-1,w}$) as the number of $+1$-labeled ($-1$-labeled) neighbors $s$ of node $i$ with edge type of $(i, s)$ in $Bin_w$. Note that $d_{i,+1} = \sum_{w=1}^{w=|W|}d_{i,+1,w}$. We then have

   	 \begin{align*}
    	\hat{P}(y_{train}=+1) = \frac{n_{+1}}{n_{+1} + n_{-1}},\\
    	\hat{P}(X_i = w | y_{train}=+1) = \frac{d_{i,+1,w} + 1}{n_{+1} + |W|};\\
    	\hat{P}(y_{train}=-1) = \frac{n_{-1}}{n_{+1} + n_{-1}},\\
    	\hat{P}(X_i = w | y_{train}=-1) = \frac{d_{i,-1,w} + 1}{n_{-1} + |W|}.
   	 %\hat{P}(X_i = 0 | y_{train}=+1) &=& \frac{n_{+1} - d_{i,+1} + 1}{n_{+1} + |W|}\\ 
   	 %\hat{P}(X_i = 0 | y_{train}=-1) &=& \frac{n_{-1} - d_{i,-1} + 1}{n_{-1} + |W|}
   	 \end{align*}

For the log-likelihood ratio, we have the following formula (see the full derivation in the  ``Derivation of Version II of Naive Bayes Classifier'' appendix section):
\begin{equation*}
log(LR(x)) \approx  \sum_{i=1}^N \sum_{w = 1}^{|W|} \mathbbm{1}_{x_i = w} \cdot log\left[ \frac{d_{i,+1,w}+1}{d_{i,-1,w}+1} \cdot \frac{n_{-1}+|W|}{n_{+1}+|W|}  \right]
\end{equation*}
	
\noindent Just like in Version I, we can make the model more flexible, by learning  different weights $\lambda_w$ for the edge bins via a logistic regression model:
\begin{align*}
\label{log-LR-weighted}
&f(\alpha, \vec{\lambda}, x) = \alpha + 
\sum_{w = 1}^{|W|} \lambda_w \underbrace{ \sum_{i=1}^N \mathbbm{1}_{x_i = w} \cdot \underbrace{log\left[ \frac{d_{i,+1,w}+1}{d_{i,-1,w}+1} \cdot \frac{n_{-1}+|W|}{n_{+1}+|W|}  \right]}_\text{one-hop aggregation by edge bin}}_\text{two-hop aggregation by edge bin}
\end{align*}

\noindent Intuitively, Version I only uses the information of the direction and weights in turning $\lambda_w$'s, while Version II also uses this edge information to estimate the posterior distribution of $y$. An alternative way for creating this classifier is to separate out one-hop and two-hop edge bins via Version II$^{\ast}$. For example, for a given node $i$, we could construct the two-hop aggregations based on the out-degree edge bin but allow different weights between a node's one-hop in-degree, out-degree, and reciprocal edges. Otherwise, Version II, would only allow out-degree aggregations among out-degree connections. We then learn different weights $\lambda_{w,w'}$ via the following:

\begin{align*}
&f(\alpha, \vec{\lambda}, x)\nonumber =\alpha +  \sum_{w = 1}^{|W|} \sum_{w' = 1}^{|W|} \lambda_{w,w'} \underbrace{ \sum_{i=1}^N \mathbbm{1}_{x_i = w} \cdot \underbrace{log\left[ \frac{d_{i,+1,w'}+1}{d_{i,-1,w'}+1} \cdot \frac{n_{-1}+|W|}{n_{+1}+|W|}  \right]}_\text{one-hop aggregation by edge bin}}_\text{two-hop aggregation by edge bin}
\end{align*}

\subsection{Empirical Results}
\textbf{Application on Facebook: }We first compare the predictive performance of the extended LINK-Naive Bayes classifier (Version I and Version II\footnote{We only include Version II$^\ast$ results for the \texttt{ogbn-proteins}.}) using the largest user-to-user interaction network -- the \emph{comment} network from Facebook. Specifically, we construct a directed, weighted user-to-user network of over 100M users based on users' commenting activities in the last week of July 2022. A directed edge will exist if a source user commented on a target user's content in the last week of July 2022, and we create edge weights based on the frequency of interactions between users. For binary prediction task, we consider the users who were exposed to Retail ads on 2022-07-30 and code $y_{train}=+1$ if a user clicked on an ad and $y_{train}=-1$ otherwise following previous ad click prediction problems~\cite{he2014practical}. We restrict this sample to active U.S.~users who are older than 18 years-old with at least 20 Facebook friends.

We set-up a within-network task where we randomly select $80\%$ nodes as training nodes, $10\%$ as validation, and the remaining $10\%$ as test nodes. Specifically, we first randomly select $80\%$ nodes where we observe their labels and merge these training labels into the edgelist. Among the rest $20\%$ nodes, we use $1/2$ to train the weights in the logistic regressions and use the remaining $1/2$ as the test nodes. Note that we don't train the weights using the $80\%$ nodes, since the NB estimators utilize the label information of the $80\%$ nodes in training, which makes them different from the remaining $20\%$ nodes. 

We report predictive performance results with normalized entropy (NE) or normalized cross-entropy~\cite{he2014practical}, which is standard for ad click prediction tasks. NE computes the predictive log loss normalized by the entropy of the baseline click-through-rate, and lower values are better. (See the ``Formula for Normalized Entropy'' appendix section for more details.) We report the relative percent change in test normalized entropy (NE) when comparing the different versions to the undirected, unweighted version in Table \ref{NB_edge_compare}. We do not report raw NE due to the sensitivity of this prediction task but all results are above a random classifier. Also note that class imbalance is not an issue for the NE metric since it uses the log-likelihood ratio, which doesn't directly measure the performance of a predictor under an imbalanced dataset. (In the next section, we'll report raw predictive results on a publicly available dataset). To test the usefulness of edge direction and weight, we consider the following three ways of dividing the edges into mutually exclusive bins:
\begin{itemize}
    \item \textbf{undirected, weighted}: we ignore the edge direction, and divide the edges by considering if the weight is of $0, 1, 2-5, >5$. We select these bins in a way to distinguish between infrequent vs. more frequent interactions. In practice, a researcher will have to select edge bins based on the empirical distribution of edge weights in their dataset or by other heuristics. 
    \item \textbf{directed, unweighted}: we treat all the edge weights the same, and divide the edges into four bins by their directions. That is, we distinguish whether the edge is only an outgoing edge from the target node, only an incoming edge from the target node, both an outgoing and incoming edge, or no edge between them.
    \item \textbf{directed, weighted}: we divide all the edges by their weight values and directions. Specifically, we consider if the out-going edge weight is of $0, 1, 2-5, >5$ and if the in-coming edge weight is of $0, 1, 2-5, >5$. There are $16$ bins in total.
\end{itemize}

\noindent Across one train/test split, we observe in Table \ref{NB_edge_compare} that the performance of Version I is better than Version II. Considering the edge direction and weights using the method introduced in Version I both seem to help improve the NE compared to the undirected/unweighted version. (Note lower NE means better performance.) We observe that considering the edge direction reduces NE by $0.09\%$ while considering the edge weight reduces NE by $0.01\%$.

\begin{table}[h]
\centering
\begin{tabular}{|c|c|c|}\hline
undirected, unweighted &\multicolumn{2}{c|}{$100\%$ (reference)}\\
\hline
{Version I} & undirected, weighted  & $-0.01\%$   \\
                        \cline{2-3}
                      & directed, unweighted  &  $-0.09\%$  \\
                      \cline{2-3}
                      & directed, weighted & $-0.11\%$  \\
\hline
{Version II} & undirected, weighted  & $+0.02\%$  \\
                        \cline{2-3}
                      & directed, unweighted  & $-0.02\%$   \\
                      \cline{2-3}
                      & directed, weighted & $-0.06\%$  \\
\hline
friend's conversion rate & \multicolumn{2}{c|}{$+0.03\%$}\\
\hline
user embeddings (PBG) & \multicolumn{2}{c|}{$+0.02\%$}\\
\hline
\end{tabular}
  \caption{Undirected/unweighted comment networks have variability in predictive performance. We observe overall higher performance with Version I compared to Version II. Both Versions exhibit higher performance than friend conversion and user embeddings from PBG.}
      \label{NB_edge_compare}
\end{table}
\noindent To compare the LINK-Naive Bayes approach with other node embedding methods, we also consider the following two baseline models:
\begin{itemize}
    \item \textbf{friend's conversion rate}: for each user $u$, we take the ratio of conversion rate among $u$'s undirected neighborhoods as the $u$'s conversion probability. We can view this as a homophily-based feature.
    \item \textbf{user embeddings (PBG)}: we implement PyTorch-BigGraph (PBG)~\cite{lerer2019pytorch} to train a 64-dimensional embedding from the user-to-user comment network. PBG scales to large graphs and can support multi-relation graphs. PBG employs a block decomposition of the adjacency matrix and trains on edges from one bucket at a time. We implement the embeddings with a 0.01 learning rate and a negative batch size of 50.
\end{itemize}

\noindent\textbf{Application on \texttt{obgn-proteins}: } We next demonstrate the LINK-Naive Bayes approach on the Node Property Prediction task of the publicly available on the Open Graph Benchmark (OGBN). OGBN includes several large-scale datasets for node prediction among other tasks and includes data splits for training models~\cite{hu2020open}. 

We include results on the \texttt{obgn-proteins} dataset, which is an undirected and weighted network consisting of 132,534 nodes, 39,561,252 edges, and 112 different labels to predict. Nodes represent proteins and edges indicate a biological association between proteins. Each edge is associated with an 8-dimensional feature representing the strength of a relationship between two proteins. This is the only binary task dataset in OGBN and the only one that allows us to set-up multi-layer networks to compare with in the next section. For this section, we will treat the network as a single layer by averaging edge weights across the 8-dimensional feature vector. For both Version I/II, we construct 4 edge bins based on intervals constructed from the 25th, 50th, 75th percentiles of the edge weight. Here we only construct edge bins based on edges that exist between nodes in order to accelerate calculations. Consistent with the data splitting recommendation~\cite{hu2020open} and with reporting ROC-AUC metrics, we split the nodes into train/validate/test splits according to what's provided and report average ROC-AUC across the 112 tasks in Table\ref{tab:protein_single_layer}. We do not incorporate any additional benchmarks as the public leaderboard already tracks top submissions.

\begin{table}[]
\centering
\begin{tabular}{|c|c|}
\hline
Method                            & Avg. ROC-AUC \\ \hline
undirected, unweighted (reference)           & 0.51         \\ \hline
Version I (undirected, weighted)  & 0.65         \\ \hline
Version II (undirected, weighted) & 0.57 \\ \hline
Version II$^{\ast}$ (undirected, weighted) &  
0.70 \\ \hline
\end{tabular}
\caption{Compare the predictive performance on the \texttt{ogbn-proteins} dataset with the LINK-Naive Bayes method using Version I/II as well as an undirected, unweighted version as a reference. We observe meaningful performance improvement in both Versions I and II.}
\label{tab:protein_single_layer}
\end{table}

For all performance metrics for this dataset, we consider ROC-AUC metrics to be consistent with results presented on the public leaderboard~\footnote{\url{ https://ogb.stanford.edu/docs/leader_nodeprop/}}. We see that Version I/II/II$^\ast$ outperform the undirected, unweighted version. We use a paired Wilcoxon signed rank test (with a Bonferroni correction) to test whether the ROC-AUC metrics from Version I/II/II$^\ast$ are greater than the undirected, unweighted version and observe $p$-values $<$ 0.01. We also test for pairwise differences between the different versions and observe a statistically significant difference. When comparing this performance to the public leaderboard as of May 2023, we can see that both Version I/II$^\ast$ are comparable to node2vec's test performance of 0.6881 at \#22 on the leaderboard. 

While we observe increased average ROC-AUC performance across the different versions relative to an undirected and unweighted version of the \texttt{ogbn-proteins} dataset, it is possible that LINK may still have more limited performance relative to the leaderboard due the train/test splits being based on species. There are more within-species edges relative to between-species edges. In the hypothetical case where there are no edges between species, the across-layer task would look more similar to an across-network task and LINK would not be applicable. 

\noindent\textbf{Summary of Empirical Results: }When comparing predictive performance for both datasets, our takeaways are that:
\begin{itemize}
    \item Adding the edge direction and/or weight can help improve predictive performance compared to an undirected/unweighted network. We observe edge weight is helpful for \texttt{ogbn-proteins} but not ad click prediction.
    \item Version I has better predictive performance than Version II for the ad click prediction task but the reverse is true for the \texttt{ogbn-proteins}. 
    \item Both sets of empirical results should be interpreted as a case study of this methodology. Whether one method is better than another will also depend in part on the underlying data generating mechanism and the network itself. 
\end{itemize}

\section{Multi-layer Networks}
\label{sec:prediction_multi}
As we previously observed in the ``Analysis of Multilayer Networks'' section, users can exhibit different interaction patterns across platforms. Such differences could potentially result in varying predictive information when using the same approach but different network layers. As a quick illustration, in Table \ref{NB_interaction_compare}, we compute the percent change in test NE using Version I (directed/weighted) for the ad click prediction task for five Facebook user-to-user interaction networks -- \emph{comment}, \emph{tag}, \emph{reshare}, \emph{mention}, and \emph{direct post}. These edges represent different types of interactions that can occur between users. A comment edge indicates a user commented on another user's content. A tag edge indicates a user tags another user on a post. This is similar to a mention but mentions can happen in comments or replies. Finally, a reshare means an edge exists if a user shares the same content from another user. We observe heterogeneity in the predictive performance using these different interaction networks.

\begin{table}[h]
  \centering
  \begin{tabular}{|c|c|c|c|c|}
  \hline 
   \emph{comment} & \emph{mention} & \emph{reshare} & \emph{tag} & \emph{post} \\
      \hline 
 $100\%$ (ref) & $+0.10\%$ & $+0.13\%$ &$+0.20\%$ &$+0.15\%$\\
%\hline
%$\#$ evaluation samples & 2684302 & 92575 & 392248 & 1954243 & 204149 \\
%\hline
%true label ratio in the evaluation & 0.066119 & 0.064834 & 0.066410 & 0.066462 &0.072413 \\
\hline
  \end{tabular}
  \caption{We compare the relative NE performance across different interaction types using the comment network as the reference network. We observe variability in predictive performance. }
      \label{NB_interaction_compare}
\end{table}
While the comment network has the best predictive performance, there is still some predictive signal in other network layers that's above a random classifier, albeit lower than just the comment network. Therefore, the question is whether we can achieve higher performance when \emph{combining} networks than any single network layer in isolation. All the above differences motivate us to further extend the LINK-Naive Bayes approach for multi-layer networks, in which we can consider multiple user-to-user networks simultaneously during training. We call this new approach MultiLayerLINK-NaiveBayes. We show in Figure~\ref{fig:single_layer_viz} (right) a schematic of combining multi-layers for a within-layer task. 

\subsection{Model for Multi-layer Networks}
We assume a $K$-layer network $\mathcal{G} = \left(\bigcup_{k=1}^K V^{(k)}, \bigcup_{k=1}^K E^{(k)}\right)$ contains a set of $k$ single-layer networks $G^{(1)} = (V^{(1)}, E^{(1)}), \cdots, G^{(K)} = (V^{(K)}, E^{(K)})$, where $V^{(k)}$ and $E^{(k)}$ represent the node set and edges representing different types of edges in network $i$, respectively. For the most comprehensive case, we consider $E^{(k)}$ to encompass both directed and weighted edges. Now, suppose we want to have the same prediction task as before, where each node in $\mathcal{V}:=\bigcup_{k=1}^K V^{(k)}$ has a label being either $+1$ or $-1$, and for a target node $u$, we want to predict its label using known labels from neighbors of $u$ at different layers.

\subsubsection{(Version 0) Linear Combination of Single layers:}
The simplest approach is to treat the multi-layer score as a linear combination of the classifiers built on each layer separately. Specifically, let $n_{+1}^{(k)}$ ($n_{-1}^{(k)}$) denote the number of $+1$-labeled ($-1$-labeled) nodes in the $k$th layer among training samples, and $d_{i,+1}^{(k)}$ ($d_{i,-1}^{(k)}$) as node $i$'s degree with $+1$ ($-1$) nodes in the $k$th layer among training samples. We similarly represent the set of bins as $\{Bin_{w^{(k)}}, w^{(k)}\in W^{(k)}\}$. Suppose layer $k$ has $N^{(k)}$ nodes, and they are divided into $|W^{(k)}|$ bins in the corresponding single-layer prediction task. We denote $\mathbbm{1}_{x^{(k)}_i = w^{(k)}}$ if the edge between the target node and the source node $i$ is of $Bin_{w^{(k)}}$ at layer $k$. For each bin in layer $k$, we set a weight $\lambda_{w^{(k)}}$, and define $\vec{\lambda} =\cup_{k=1}^K \cup_{w^{(k)}=1}^{|W^{(k)}|} \lambda_{w^{(k)}}$. Following Equation (\ref{log-LR-weighted-v1}), we can write the score function as:

\begin{align}
&f(\alpha, \vec{\lambda}, x)\nonumber = \alpha + \sum_{k=1}^K \sum_{w^{(k)} = 1}^{|W_k|} \lambda_{w^{(k)}} \underbrace{\sum_{i=1}^{N^{(k)}} \mathbbm{1}_{x^{(k)}_i = w^{(k)}}  \underbrace{log\left[ \frac{d_{i,+1}^{(k)}+1}{d_{i,-1}^{(k)}+1} \cdot \frac{n_{-1}^{(k)}+2}{n_{+1}^{(k)}+2} \right]}_\text{one-hop aggregation at layer $k$}}_\text{two-hop aggregation by edge bin at layer $k$}
\end{align}
This score function is a linear combination of the scores from different layers. As before, we can then estimate $\alpha$ and 
$\vec{\lambda}$ with Logistic regression to minimize the log loss. 

% \begin{eqnarray}
%     l := \sum_{(X, y) \in {\text{training data}}} -\left(\frac{y+1}{2}\right) \log\left(\frac{e^{f(\alpha, \vec{\lambda}, X)}}{1+e^{f(\alpha, \vec{\lambda}, X)}}\right) \nonumber \\
%          - \left(\frac{y-1}{2}\right) \log\left(\frac{1}{1+e^{f(\alpha, \vec{\lambda}, X)}}\right).
% \end{eqnarray}

\subsubsection{(Version I) Learning Weights for Edge Bins:} Alternatively, we can couple edges between layers and place them into edge bins based on the product category of bins at all single layers. That is, we first  construct edge bins based on edge weight or direction 
%ordered pairs of nodes in multi-layer networks into $\tilde{m}$ bins by considering whether there is an edge between the ordered pair at layer $k$, as well as the edge weight and direction at layer $k$. For example, at each layer, node pairs are first divided by whether they are adjacent. They can then be divided by their edge weight values or quantiles and by whether their edge directions 
as introduced in the ``Create Edge Bins'' subsection. For a given edge that exists across multiple network layers, we categorize it into one of the bins based on edge weight/direction for that layer. Then to couple an edge across multiple layers, we consider the joint categorization of edge bins across multiple layers. The final bins in the multi-layer network are the product category of bins at all single layers. That is, if at each layer $k$, the edges are divided into $|W^{(k)}|$ bins, then the total number of bins for the $K$-layer network is $\prod_{k=1}^K |W^{(k)}|$. As before, we denote the set of bins as $\{Bin_{\tilde{w}}, \tilde{w} \in \tilde{W} \}$, where $\tilde{W}$ is the set of bins' labels, and we denote $\mathbbm{1}_{x_i = \tilde{w}}$ if the relationship between the target node and the source node $i$ is of $Bin_{\tilde{w}}$. 

We then assign different weights for different bins generated at different layers. Let $\lambda_{\tilde{w}}^{(k)}$ be the weight of $Bin_{\tilde{w}}$ for layer $k$, and denote by $\vec{\lambda}$ the vector of $\lambda_{\tilde{w}}^{(k)}$'s: $\vec{\lambda} = \cup_{k=1}^K \cup_{\tilde{w}=1}^{|\tilde{W}|} \lambda_{\tilde{w}}^{(k)}$. We can write the classifier's score function as the following:

\begin{align*}
&f(\alpha, \vec{\lambda}, x)=\alpha  + \sum_{k=1}^K\sum_{\tilde{w} = 1}^{|\tilde{W}|} \lambda_{\tilde{w}}^{(k)} \underbrace{\sum_{i=1}^N \mathbbm{1}_{x_i = \tilde{w}} \cdot \underbrace{log\left[ \frac{d_{i,+1}^{(k)}+1}{d_{i,-1}^{(k)}+1} \cdot \frac{n_{-1}^{(k)}+2}{n_{+1}^{(k)}+2} \right]}_\text{one-hop aggregation at layer $k$}}_\text{two-hop aggregation by coupling edge bins across layers}
\end{align*}

As explained, in this method we construct the finer bins $\{Bin_{\tilde{w}}, \tilde{w} \in \tilde{W} \}$ by the Cartesian product of each layer' bins. We then weight the scores in each layer using the finer bins, and the score function is a linear combination of every layers' scores. The advantage of using the finer bins is that it contains multi-layer information of a node, which might capture some interaction effect across different layers, that cannot be captured by Version 0 above. 

\subsubsection{(Version II/II$^\ast$) Classifier within each Edge Bin:} Similar to the Version II in the single-layer network section, we can apply a Naive Bayes classifier within each bin for multi-layer networks. We denote $d_{i,+1,\tilde{w}}$ ($d_{i,-1,\tilde{w}}$) as the number of $+1$-labeled ($-1$-labeled) neighbors $s$ of node $i$ with edge type of $(i, s)$ in $Bin_{\tilde{w}}$. We also denote by $n_{-1}$ ($n_{+1}$) the number of nodes in $\mathcal{V}$ with the label $-1$ ($+1$).

%Note that $d_{i,+1} = \sum_{w=1}^{w=|W|}d_{i,+1,w}$. 

With exactly the same derivation as Version II in in the previous single-layer section, we have the log-likelihood ratio as follows:
\begin{eqnarray*}
log(LR(x)) \approx \sum_{i=1}^N \sum_{\tilde{w} = 1}^{|\tilde{W}|} \mathbbm{1}_{x_i = \tilde{w}} \cdot log\left[ \frac{d_{i,+1,\tilde{w}}+1}{d_{i,-1,w}+1} \cdot \frac{n_{-1}+|\tilde{W}|}{n_{+1}+|\tilde{W}|}  \right]
\end{eqnarray*}
	
\noindent Again, we can set different weights $\lambda_{\tilde{w}}$ on different bins and train the weights with Logistic regression to minimize the log-loss.
\begin{align*}
f(\alpha, \vec{\lambda}, x) &=\alpha +\sum_{\tilde{w} = 1}^{|\tilde{W}|} \lambda_{\tilde{w}} \sum_{i=1}^N \mathbbm{1}_{x_i = \tilde{w}} \cdot log\left[ \frac{d_{i,+1,\tilde{w}}+1}{d_{i,-1,\tilde{w}}+1} \cdot \frac{n_{-1}+|\tilde{W}|}{n_{+1}+|\tilde{W}|} \right]
\end{align*}

\noindent Intuitively, in this version we still construct the finer bins $\{Bin_{\tilde{w}}, \tilde{w} \in \tilde{W} \}$ by the Cartesian product of each layer' bins. In Version I we only use the bins division to weight the one-hop scores in different layers, while in Version II we calculate the one-hop score directly using the finer bins. To create Version II$\ast$, we similarly separate out the one-hop and two-hop edge bins as in the single-layer set-up.

\subsection{Empirical Results}
\textbf{Application on Facebook: }  To test the performance of the multi-layer LINK-Naive Bayes classifier, we use the two largest Facebook interaction networks -- \emph{comment} and \emph{mention} -- to construct a two-layer network. As before, we select active U.S.~users with at least 20 Facebook friends who were exposed to Retail ads on 2022-07-30. The number of edge bins in the two-layer network is the product of the number of bins for each network layer. Since our single-layer results comparing edge direction and weight show that only edge direction improves predictive performance across both versions in Table~\ref{NB_edge_compare}, for simplicity, we ignore the edge weights and we build directed, unweighted user-to-user networks based on uses' commenting or mentioning activities. Within each layer, for a directed pair of nodes $(u,v)$, there are four possible relationships between them: there is no edge between $u$ and $v$, there is a directed edge going from $u$ to $v$, there is a directed edge going from $v$ to $u$, and there are reciprocal edges. 

We then set-up a within-network task similar to the within-network section where we randomly select $80\%$ nodes as the training sample, $10\%$ as validation sample, and the remaining $10\%$ as test sample. We report the percent change in test NE in Table~\ref{tab:multi_layer_comparison} relative to the baseline of using the single-layer comment network.
\begin{table}[h]
    \centering
    \begin{tabular}{|c|c|}
        \hline
        comment (Version I in Sec 3) & $100\%$ (reference) \\
        \hline
        mention (Version I in Sec 3) & $+0.09\%$ \\
        \hline
        comment + mention (Version 0) & $-0.03\%$\\
        \hline
        comment + mention (Version I) & $-0.07\%$\\
        \hline
        comment + mention (Version II) & $+0.01\%$ \\
        \hline
        comment + mention (PBG) & $+0.06\%$ \\
        \hline
    \end{tabular}
    \caption{We compare MultiLayerLINK-NaiveBayes across the different Versions 0/I/II and compare to the single-layer network prediction performance. All NE metrics are compared relative to the single-layer network (comment for Version I). }
    \label{tab:multi_layer_comparison}
\end{table}

\noindent We observe that:
\begin{itemize}
    \item For Version 0: adding the multi-layer information (from the mention network) is helpful compared to just the comment network in isolation. 
    \item For the multi-layer Version I approach: it outperforms Version 0. We expect this since Version I uses the finer bins (Cartesian product of each layer's bins) to weight the scores, which can capture the interaction effect across different layers that is missing in Version 0.
    \item For Version II: it performs even worse than the single layer baseline. One explanation could be that, our data set is imbalanced (only about 6 percent of nodes have label 1), so when we calculate the Naive Bayes scores directly for each bin (which might only contain a small number of data points), the estimation of the posterior probability might be highly volatile.
\end{itemize}

\noindent\textbf{Application on \texttt{obgn-proteins}: } We next compare predictive performance on the \texttt{obgn-proteins} described in the earlier single-layer section. We create multi-layer networks by utilizing the 8-dimensional edge weight vector to construct 8 separate network layers corresponding to different edge weights. Within each layer, for an undirected pair of nodes $(u,v)$, we bin edges based on a 75th percentile cutoff for edge weights and only consider bins where an edge exists. We again use the same train/validation/test splits described in the earlier section and that's available in the public data. We report the Average ROC-AUC in Table~\ref{tab:protein_multi_layer}. We observe that:
\begin{itemize}
    \item For Version 0, we observe that Version I/II/II$\ast$ have higher average predictive performance.
    \item For the multi-layer set-up, Version I/II/II$^\ast$ both have higher average ROC-AUC compared to their single-layer counterparts with $p$-values $<$0.01.
    \item We do not observe a meaningful performance difference in the multi-layer setting for Version II vs. II$^\ast$.
\end{itemize}

\begin{table}[h!]
\centering
\begin{tabular}{|c|c|}
\hline
Method                            & Avg. ROC-AUC \\ \hline
Version 0 (undirected, weighted)           &      0.68    \\ \hline
Version I (undirected, weighted)  &      0.72    \\ \hline
Version II (undirected, weighted) &   0.75      \\ \hline
Version II$^{\ast}$ (undirected, weighted) &  
0.76 \\ \hline
\end{tabular}
\caption{Compare the predictive performance on the \texttt{ogbn-proteins} dataset with the MultilayerLINK-Naive Bayes method using Version 0/I/II/II$^\ast$ for the multi-layer network set-up.}
\label{tab:protein_multi_layer}
\end{table}

\subsection{Time Complexity of MultiLayerLINK-NaiveBayes}
We finally compute the time complexity of the multilayer approach. 
\label{appendix-time-complexity}
For a $k$-layer network $\mathcal{G} = (\mathcal{V}, \mathcal{E})$ with $\mathcal{V} = \cup_{k=1}^K V^{(k)}$, and $\mathcal{E}= \cup_{k=1}^K E^{(k)}$, assume that we have $l$ bins; that is, $\vec{\lambda}$ is of dimension $l$ in the logistic regression step. The time complexities for different steps are:
\begin{itemize}
\item for getting the NB estimators, the time complexity is $O(\vert\mathcal{E} \vert )$,
\item for calculating the score functions for all nodes, the time complexity is $O(\vert \mathcal{E} \vert)$, and
\item for training the logistic regression, the time complexity is $O(l \cdot\vert  \mathcal{V} \vert)$.
\end{itemize}
In total, the complexity for our model is $O(\vert \mathcal{E} \vert + l \cdot \vert \mathcal{V} \vert)$. 

\section{Discussion and Future Work}
\label{sec:conclude}
Social network platforms enable users to connect with one another in numerous ways. As a result, organizations have an opportunity to combine and learn from different user-to-user network representations. In this paper, we first demonstrate that the various IG/FB user-to-user networks exhibit different properties and minimal edge overlap. Given these empirical differences between user networks, we first extend a single-layer network prediction approach called LINK-Naive Bayes to account for edge direction and weights. We introduce a business-relevant prediction task, ad click prediction, as well as utilize publicly available data from the Open Graph Benchmark to compare our different approaches. We observe the most performance gain for ad click prediction in incorporating edge direction. For the Open Graph Benchmark dataset, ROC-AUC results improve from 0.51 in the undirected and unweighted case to 0.57-0.70 in the different versions for incorporating weight information. Next, we introduce MultiLayerLINK-NaiveBayes, which incorporates multi-layer network signals for the within-network prediction task. We observe higher predictive performance gain for Version I for ad click prediction and Version I/II/II$\ast$ relative to their single-layer counterparts. The results in our paper should be viewed as a case study on how to deploy LINK to multi-layered networks. We may see in other settings that other versions do better than what was presented in this work. 

Future work could consider comparing approaches on synthetic graphs~\cite{tsitsulin2022synthetic} for generalizing when this method does better to alternatives. While the results in this paper were shown for binary classification and multiplex networks, the methodology can generalize to multi-class prediction and more general multilayer networks. For multi-class prediction, we could extend by treating it as a one-vs-rest binary task across each class. For more general multilayer networks, we could extend the methodology by creating edge bins based on how nodes are connected across layers. We leave these extensions for future work.  Beyond node prediction, LINK-NB is applicable to link prediction as discussed in the ``Link Prediction'' appendix section.

We also consider additional possible directions for extending our work:

\begin{itemize}
    \item \emph{Temporal networks}: Although our study focuses on predictions on network representations at a particular time point, user networks are dynamic over time. If networks are correlated over time, then we'd expect our approach to generalize to prediction over time. However, if networks are very different across time, then future work would need to account for the changes in edges over time. One option is to encode temporal motifs~\cite{paranjape2017motifs} or to account for time intervals between new edges appearing.
    \item \emph{Missingness mechanisms and noisy observations}: In our current study, we assume node labels are missing completely at random (MCAR)~\cite{little2019statistical}. Across various business prediction tasks, the MCAR assumption is likely unreasonable. For example, for demographic user prediction~\footnote{\small{\url{https://tech.facebook.com/artificial-intelligence/2022/06/adult-classifier/}}} we may expect training data labels to vary by other characteristics. In these cases, further work is needed to understand how robust our approach is to violations of MCAR. We also assume we observe accurate ground-truth of node labels and edges. GNN can under perform when there are noisy edges~\cite{liu2022towards} and future work may consider how robust LINK and its variants are to noise.
    \item \emph{Cold start for new users}: Finally, we assume the set of users in the sample is known and fixed. However, in practice, new users are joining interactions over time. We consider new users a cold start problem because there aren't existing network signals to leverage. One thought is trying to identify similar nodes in the network and using those as a proxy.
\end{itemize}

In total, combining multi-layers network signals and accounting for both edge direction and weights for with-network prediction is a critical step forward in enabling organizations to optimally learn from network representations.\newline

\noindent\textbf{Code Availability: } The code to replicate the analysis on the \texttt{ogbn-proteins} is available at \url{https://github.com/facebookresearch/LINKExtension}.

%%
%% The next two lines define the bibliography style to be used, and
%% the bibliography file.

\bibliographystyle{unsrt}
\bibliography{ms.bib}

%%
%% If your work has an appendix, this is the place to put it.
\appendix

\section{Appendix}

\subsection{Derivation of Version II of Naive Bayes Classifier}
\label{appendix-derivation-II}
Recall that we have the estimate
\begin{align*}
    	& \hat{P}(y_{train}=+1) = \frac{n_{+1}}{n_{+1} + n_{-1}},\\
    	&  \hat{P}(X_i = w | y_{train}=+1) = \frac{d_{i,+1,w} + 1}{n_{+1} + |W|};\\
    	&\hat{P}(y_{train}=-1) = \frac{n_{-1}}{n_{+1} + n_{-1}},\\
    	&\hat{P}(X_i = w | y_{train}=-1) = \frac{d_{i,-1,w} + 1}{n_{-1} + |W|}.
\end{align*}
   	 
By the formula of the Naive Bayes classifier (conditional independence), we have that
    \begin{align*}
	LR(x) &= \frac{P(y_{train}=+1)}{P(y_{train}=-1)}
	\cdot \prod_{i=1}^N \frac{P(X_i = x_i|y_{train}=+1)}{P(X_i = x_i|y_{train}=-1)} \\
	&= \frac{P(y_{train}=+1)}{P(y_{train}=-1)}
	\cdot \prod_{w=1}^{|W|} \prod_{i: x_i = w} \frac{P(X_i = w|y_{train}=+1)}{P(X_i = w|y_{train}=-1)}
    \\
    &= \frac{n_{+1}}{n_{-1}} \prod_{w=1}^{|W|} \prod_{i: x_i = w} \frac{d_{i,+1,w} + 1}{d_{i,-1,w} + 1} \cdot \frac{n_{-1} + |W|}{n_{+1} + |W|}.
    \end{align*}
Plugging in our estimates into the above equation, and taking log on both sides, we get %\ka{comment for webconf, I think reviewers are going to want to see full derivation -- we can work on when you're back}
    \begin{align*}
	log(LR(x)) &\approx C_2 +\sum_{i=1}^N \sum_{w=1}^{|W|} \mathbbm{1}_{x_i = w} log\left[ \frac{d_{i,+1,w}+1}{d_{i,-1,w}+1} \cdot \frac{n_{-1}+|W|}{n_{+1}+|W|} \right] \\
	            %&\approx C_2 + \sum_{i=1}^N \sum_{w=1}^{|W|} \mathbbm{1}_{x_i = w} log\left[ \frac{d_{i,+1,w}+1}{d_{i,-1,w}+1} \cdot \frac{n_{-1}+|W|}{n_{+1}+|W|} \right],
    \end{align*}

\noindent which gives the expression for the Version II approach.

\subsection{Generalizations of LINK-NB to Link Prediction}
\label{appendix-link-prediction}
% \ka{We will also add a proof here of how our methodology generalizes to link prediction. We will not be including any new data results but will only demonstrate how the approach can generalize to other common network problems.}

In this section we discuss how to apply LINK-NB for edge attributes prediction. Suppose that each pair of nodes $(z_1,z_2)$ in the training data has a binary label $e_{(z_1, z_2)} \in \{+1, -1\}$. For example, $e_{(z_1, z_2)}$ can be the indicator of whether $z_1$ and $z_2$ are friends. 

Our features are the rows of the adjacency matrix that corresponds to $z_1, z_2$. That is, when ignoring edge directions and edge weights, for all node $i \in G$, we define $x_i := (\mathbbm{1}_{i \text{ is a neighbor of } z_1}$, $\mathbbm{1}_{i \text{ is a neighbor of } z_2})$ as the feature vector of $i$. Note that there are four possible values for $x_i$, and we can group the nodes by their values into four bins, which we call them as $Bin_w$ for $w = 1,\ldots, 4$. In the more complicated cases where we consider edge weights or directed edges or multiple layers, similar to node LINK-NB, we can also group nodes by values of $x_i$ into generalized bins $\{Bin_w\vert  w = 1,\ldots, |W|\}$. We denote $\mathbbm{1}_{x_i = w}$ as $x_i \in Bin_w$.

We have the following expression for the likelihood ratio for the Naive Bayes estimator
\begin{align}
LR(x) &= \frac{P(e_{train}=+1)\cdot P(x | e_{train}=+1)}{P(e_{train}=-1) \cdot P(x | y_{train}=+1)} \\
&= \frac{P(e_{train}=+1)\cdot \prod_{i=1}^{N}P(x_i | e_{train}=+1)}{P(e_{train}=-1)\cdot \prod_{i=1}^{N}P(x_i | e_{train}=-1)} \label{eq_nb_link}
\end{align}

Let $n_{+1}$ and $n_{-1}$ be the count of $(z_1, z_2)$ pairs with $e_{(z_1, z_2)}=+1$ and $e_{(z_1, z_2)}=-1$ respectively. Denote by $d_{i,+1,w}$ (and $d_{i,-1,w}$) the number of $(z_1, z_2)$ pairs with label $+1$ (and $-1$), such that its corresponding $x_i = w$. Similar to previous sections, we have the following estimates,
   	 \begin{align*}
    	\hat{P}(e_{train}=+1) &= \frac{n_{+1}}{n_{+1} + n_{-1}},\\  \hat{P}(e_{train}=-1) &= \frac{n_{-1}}{n_{+1} + n_{-1}}, \\
    	\hat{P}(x_i = w | e_{train}=+1) &= \frac{d_{i,+1,w} + 1}{n_{+1} + |W|},\\ 
    	\hat{P}(x_i = w | e_{train}=-1) &= \frac{d_{i,-1,w} + 1}{n_{-1} + |W|}.
   	 \end{align*}

\noindent Substituting the above estimates into (\ref{eq_nb_link}), we get the following expression for the log likelihood ratio:
\begin{align*}
log(LR(x)) &\approx C_4 + \sum_{i=1}^N \sum_{w = 1}^{|W|} \mathbbm{1}_{x_i = w} \cdot log\left[ \frac{d_{i,+1,w}+1}{d_{i,-1,w}+1} 
                   \cdot \frac{n_{-1} + |W|}{n_{+1} + |W|} \right]
\end{align*}

\noindent Similar to the node attributes prediction case, we can then set different weights, $\lambda_j$'s, on different categories of $x_i$, and train $\alpha$ and the weights with Logistic regression with the score function
\begin{align*}
\label{log-LR-weighted}
f(\alpha, \vec{\lambda}, x) &= \alpha + \sum_{w = 1}^{|W|} \lambda _ j \sum_{i=1}^N \mathbbm{1}_{x_i = w} \cdot log\left[ \frac{d_{i,+1,w}+1}{d_{i,-1,w}+1} 
                   \cdot \frac{n_{-1} + |W|}{n_{+1} + |W|} \right]
\end{align*}

\subsection{Formula for Normalized Entropy}
\label{formula_NE}
{\color{black}Normalized entropy is the predictive log loss normalized by the entropy of the average empirical clicking rate. The lower the value is, the better predictive power the model has. Suppose there are $N$ samples in the dataset, where each sample has a label $y_i\in \{+1, -1\}$, then the normalized entropy is defined as:
\begin{equation}
    NE = \frac{-\frac{1}{N}\sum_{i=1}^n \left(\frac{1+y_i}{2}\log(p_i) + \frac{1-y_i}{2}\log(1-p_i)\right)}{-(p*\log(p) + (1-p)*\log(1-p)},
\end{equation}
where $p_i$ is the estimated probability of user $i$'s clicking rate, and $p$ is the average clicking rate in the dataset.}
\end{document}